\documentstyle[prl,aps,epsf,twocolumn]{revtex}
\begin{document}
\draft
\twocolumn[
\hsize\textwidth\columnwidth\hsize\csname @twocolumnfalse\endcsname

\title{Fluctuational phase-flip transitions in 
parametrically pumped oscillators}

\author{ M.I.~Dykman, C.M.~Maloney,  V.N.~Smelyanskiy and M.~Silverstein}

\address{Department of Physics and Astronomy, 
Michigan State University, East Lansing, MI 48824}

\date{\today}
\maketitle

\widetext
\begin{quote}
We analyze the rates of noise-induced transitions between period-two
attractors. The model investigated is an underdamped oscillator
parametrically driven by a field at nearly twice the oscillator
eigenfrequency. The activation energy of the transitions is analyzed
as a function of the frequency detuning and field amplitude scaled by
the damping and nonlinearity parameters of the oscillator. The
parameter ranges where the system is bi- and tristable are
investigated. Explicit results are obtained in the limit of small
damping (or strong driving), and near bifurcation points.

\end{quote}

]
\narrowtext
\section{Introduction}

Nonlinear systems driven by a sufficiently strong periodic field often
display period doubling \cite{Lichtenberg}. The two emerging stable
periodic states are the same: the only difference between them is that
they are shifted in time by the period of the force
$2\pi/\omega_F$. This feature is a consequence of the symmetry with respect to
translation in time by $2\pi/\omega_F$, and it has attracted much
attention to such systems as elements of digital computers. In the
presence of noise, there occur fluctuational transitions between the
period-two attractors, which correspond to phase slip of the system by
$\pi$.  Such transitions have been observed recently for electrons
oscillating in a Penning trap \cite{Tan}. They have been also
investigated in an analog electronic circuit \cite{Carroll}, in the
context of stochastic resonance \cite{SR}.

Motivated by these observations, in the present paper we develop a
theory of escape rates from period-two attractors. The analysis is
done for the simplest generic model that displays period doubling: an
underdamped oscillator parametrically driven by a force at nearly
twice the oscillator eigenfrequency $\omega_0$
\cite{Mechanics}. This model applies, in particular, to axial
vibrations of an electron in a Penning trap \cite{Tan}.  Much work on
a parametrically excited oscillator has been done in the context of
squeezed states of light, cf. \cite{squeezing}. We analyze escape due
to classical fluctuations, which were substantial for the systems
investigated in \cite{Tan,Carroll}.

Parametrically excited oscillator is an example of a system away
from thermal equilibrium. Such systems usually lack detailed balance
\cite{Landauer}, they are not characterized by free energy, and escape
rates depend on the system dynamics and the noise that gives rise to
fluctuations in the system. In the important and quite general case
where the noise is Gaussian, there has been developed a technique
which reduces the problem of calculating escape rates to a variational
problem \cite{Lindenberg}. The solution of this problem describes the
optimal path along which the fluctuating system is most likely to move
when it escapes. Such path is often called the most probable escape
path (MPEP) \cite{Maier}. The value of the variational functional
gives the exponent in the expression for the escape rate.

An advantageous feature of the problem of period doubling in an
underdamped oscillator is that the period doubling occurs for
comparatively small amplitudes $F$ of the driving force, where the
nonlinearity of the oscillator is still small: the anharmonic part of
the potential energy is much less than the quadratic term
$\omega_0^2q^2/2$. In this case, the quantities of interest are the
amplitude and phase of the vibrations at the frequency
$\omega_F/2\approx \omega_0$. They vary only a little over the time
$\sim \omega_F^{-1}$. The corresponding dynamics is affected by
Fourier components of the noise within a narrow band centered at
$\omega_F/2$. Essentially, this means that, in the analysis of the
dynamics of slow variables, the noise may be assumed white. A similar
situation arises
\cite{Krivoglaz} in 
the problem of transitions between the stable states of forced
vibrations of a resonantly driven underdamped oscillator.

Below in Sec.~II we provide the results on the the phase portrait of
the oscillator in the rotating frame, derive the properties of noise
for slow variables, and, for low noise intensities, formulate the
variational problem for the activation energy of escape. In Sec.~III
we provide explicit expressions for the escape rates in the following
parameter ranges: vicinities of the bifurcation points where period
two attractors emerge and where there emerge two unstable period two
states, and the range of comparatively strong driving where the motion
in {\it slow} variables is underdamped. In Sec.~IV numerical results
for the general case are discussed and compared with the results for
the limiting cases. Sec.~V contains concluding remarks.

\section{Asymptotic expression for the escape rates}

\subsection{Phase portrait in slow variables}

To set the scene, we will first discuss the phase portrait of a
parametrically driven underdamped oscillator. A simple
phenomenological equation of motion is of the form

\begin{equation}
{d^2q\over dt^2} + 2\Gamma {dq\over dt}+ \omega_0^2q + \gamma q^3+
qF\cos\omega_F t = \xi(t)\label{ddot}
\end{equation}

\noindent
Here, $\Gamma$ is the friction coefficient, $\gamma$ is the
nonlinearity parameter, $F$ is the amplitude of the regular force, and
$\xi(t)$ is a zero-mean noise, $\langle \xi(t)\rangle = 0$.  The
effect of cubic nonlinearity in the Hamiltonian of the oscillator can
be taken into account by renormalizing the parameter $\gamma$
(cf. \cite{Mechanics}). The results can be also easily generalized to
the case where, for special reasons \cite{Tan}, $\gamma$ is
numerically small, and it is necessary to allow for the term $\propto
q^5$ in the equation of motion.

We consider resonant driving, so that
the oscillator eigenfrequency $\omega_0$ is close to $ \omega_F/2$,

\begin{equation}
\Gamma, |2\omega_0-\omega_F| \ll \omega_0. \label{conditions}
\end{equation}

\noindent
In this case it is convenient (cf. \cite{Mechanics}) to analyze the
oscillator motion in the rotating frame. Respectively, we change to
slow dimensionless time $\tau = \Gamma t$ and slow dimensionless
variables $q_1, q_2$:

\begin{eqnarray}
&&q(t) = \left({4\omega_F\Gamma\over 3|\gamma|}\right)^{1/2}
\left[q_1 \cos{\omega_Ft\over 2} -
q_2\sin{\omega_Ft\over 2}\right], \label{slow}\\ 
&&{d q\over dt} = -\left({\omega_F^3\Gamma\over 3|\gamma|}\right)^{1/2}
\left[q_1 \sin{\omega_Ft\over 2} +
q_2\cos{\omega_Ft\over 2}\right].\nonumber
\end{eqnarray}

Following the standard procedure of the method of averaging
(cf. \cite{Lichtenberg}) and neglecting fast oscillating terms which
depend on the oscillator amplitude and contain a factor $\exp(\pm
in\omega_Ft/2)$ with $n\neq 0$, one obtains the equations of
motion for $q_1,q_2$ in the form

\begin{eqnarray}
\dot q_1 &\equiv& {dq_1\over d\tau}= -q_1 + {\partial g\over \partial
q_2} + \xi_1(\tau/\Gamma),\; \tau \equiv \Gamma t,\label{SDE}\\ 
\dot q_2 &\equiv& {dq_2\over d\tau}=
-q_2 - {\partial g\over \partial q_1} + \xi_2(\tau/\Gamma)\nonumber
\end{eqnarray}

\noindent
where $\xi_{1,2}(\tau/\Gamma)$ are the random forces, and

\begin{eqnarray}
g(q_1,q_2)&&= {1\over 2}(q_1^2+q_2^2)\left[\Omega - {1\over 2} (q_1^2
+ q_2^2){\rm sgn}\,
\gamma\right]\label{g}\\
&& + {1\over 2}\zeta (q_2^2-q_1^2).\nonumber
\end{eqnarray}

\noindent
In what follows we assume that $\gamma > 0$; the case $\gamma < 0$
can be described by replacing $\Omega\rightarrow - \Omega,\;
q_i\rightarrow -q_i \;(i=1,2)$, and by changing the phase of the field
in Eq.~(\ref{ddot}) by $\pi$.

Except for the random force, the motion of the oscillator as described
by Eqs.~(\ref{SDE}) is characterized by two dimensionless
parameters: the scaled frequency detuning $\Omega$ and the scaled
field $\zeta$,

\begin{equation}
\Omega =
[(\omega_F/2)-\omega_0]/ \Gamma, \quad \zeta = F/
2\omega_F \Gamma. \label{parameters}
\end{equation}

\noindent
For $\zeta < 1$ or for $\Omega < -(\zeta^2-1)^{1/2}$ the oscillator
(\ref{SDE}) in the absence of noise has only one stable state,
$q_1=q_2 =0$: parametric oscillations are not excited. The value $\zeta
= 1$ gives the threshold field amplitude $F_{th}= 2\omega_F$ which is
necessary for exciting period-two vibrations. The phase portrait of
the oscillator in variables $q_1,q_2$ in the range where the
oscillations are excited is shown in Fig.~1. 

For $\zeta >1$, with the increasing $\Omega$ there first occurs a
pitchfork bifurcation for $\Omega = -(\zeta^2-1)^{1/2}$. Here, the
stable state $q_1=q_2=0$ becomes unstable and there emerge two stable
states ${\bf q}_{st}^{(1,2)}$ which are symmetric with respect to
$q_1=q_2=0$, see Fig.~1a. These states correspond to stable period
two vibrations shifted in phase by $\pi$. The vibrational amplitude $a
= |{\bf q}_{st}|$ increases monotonically with the increasing $\Omega$
(see inset in Fig.~1). The results of the asymptotic analysis based on
Eqs.~(\ref{SDE}) apply for not too large amplitudes $a$,

\begin{equation}
\Gamma a^2, |2\omega_F-\omega_0|a^2 \ll \omega_F. \label{limitation}
\end{equation}

As $\Omega$ goes through the value $(\zeta^2-1)^{1/2}$ there occurs
the second pitchfork bifurcation of the state $q_1=q_2=0$: this state
becomes stable again, and there emerge two unstable states ${\bf
q}_{u}^{(1,2)}$, which are also symmetric with respect to the origin
and correspond to unstable period two vibrations of the oscillator. As
it is seen from Fig.~1b, the phase plane of the oscillator is such
that, for $\Omega > (\zeta^2-1)^{1/2}$ the domains of attraction to
the stable states ${\bf q}_{st}^{(1)}, {\bf q}_{st}^{(2)}$ are
separated by the domain of attraction to the stable state ${\bf q}=0$.

\subsection{Noise in the equations for slow variables}

In many cases of physical interest, the noise $\xi(t)$ driving the
oscillator is Gaussian. This noise may originate from the coupling to
a thermal bath (which also gives rise to the friction force in
Eq.~(\ref{ddot})), or it may be due to an external nonthermal
source. A zero-mean Gaussian noise is characterized by its power
spectrum

\begin{equation}
\Phi_{\omega}[\xi(t+t'),\xi(t)] = \phi(\omega)=
\int_{-\infty}^{\infty} dt'\,e^{i\omega t'}
\langle \xi(t+t')\xi(t)\rangle . \label{Phi}
\end{equation}

\noindent
For a stationary noise the power spectrum (\ref{Phi}) is independent
of time.  In what follows we assume that the function $\phi(\omega)$
is {\it smooth} near the oscillator eigenfrequency $\omega_0$.

The noises $\xi_{1,2}(t)$ in Eq.~(\ref{SDE}) are nonstationary,
generally speaking. One obtains from Eqs.~(\ref{ddot}), (\ref{slow}),
(\ref{SDE}) the following expressions for their power spectra:

\begin{eqnarray}
\Phi_{\omega}[\xi_i(t+t'),\xi_i(t)]
&&={3|\gamma|\over 8\omega_F^3\Gamma}\phi\left(\omega- {1\over
2}\omega_F\right)\nonumber\\
&&\times\left[1-(-1)^i\cos\omega_Ft\right] \;(i=1,2),\label{corr1}\\
\Phi_{\omega}[\xi_1(t+t'),\xi_2(t)]&&
=
-{3|\gamma|\over 8\omega_F^3\Gamma}\phi\left(\omega- {1\over
2}\omega_F\right)\sin \omega_Ft.\nonumber
\end{eqnarray}

\noindent
It follows from Eq.~(\ref{corr1}) that the spectra of the diagonal
correlators $\langle \xi_i(t)\xi_i(t')\rangle$ have both
time-independent and fast-oscillating in time components, whereas the
power spectrum of the cross-correlator of  $\xi_1, \xi_2$
is fast oscillating in time. Therefore, in the analysis of the effect
of the noise on the slowly varying functions $q_{1,2}$, in the spirit
of the averaging method, one can assume that the noise components
$\xi_1(t), \xi_2(t)$ are asymptotically independent of each other; one
can also leave out the terms $\propto \cos \omega_Ft$ in the power
spectra of the diagonal correlators (cf. \cite{Krivoglaz} where the
analysis was done for a microscopic model of noise resulting from
coupling to a bath; rigorous mathematical results on the method of
averaging in noise-driven systems are discussed in
\cite{Freidlin}). 

The dynamics of the slow variables $q_{1,2}$ is characterized by the
time scales $\sim 1/\Gamma, 1/|2\omega_F-\omega_0|$. If  the power
spectrum $\phi(\omega)$ varies only slightly in the whole range
$|\omega - \omega_F/2|\alt \Gamma, |2\omega_F-\omega_0|$ (and this
range does not correspond to a deep minimum of $\phi(\omega)$), then
one can assume that the random functions $\xi_1(t)$, $\xi_2(t)$ are
asymptotically independent zero-mean Gaussian white noises,

\begin{eqnarray}
&&\langle \xi_i(\tau/\Gamma)\xi_i(\tau'/\Gamma)\rangle \approx 
D\tilde\delta(\tau - \tau'), \;i=1,2 \label{correlator}\\
&&D = (3|\gamma|/ 8\omega_F^3\Gamma^2)\phi(\omega_F/2)\nonumber
\end{eqnarray}

\noindent
(the function $\tilde\delta(x)$ is of the type of a $\delta$-function:
it is large in the domain $|x|\ll 1$, and its integral is equal to 1).

The quantity $D$ gives the characteristic intensity of the noise in
the equations of motion for slow variables $q_1,q_2$ (\ref{SDE}). In
particular, if the oscillator is coupled to a thermal bath with the
correlation time much smaller than $1/\omega_0$, so that it performs a
``truly'' Brownian motion, and the random force $\xi(t)$ in
Eq.~(\ref{ddot}) is $\delta$-correlated, with intensity $4\Gamma kT$,
we have $D=3|\gamma|kT/ 2\omega_F^3\Gamma$. We note, however, that the
dynamics of slow variables can be described as Brownian motion even
where this description does not apply to the motion of the initial
oscillator, i.e. where the correlation time of the thermal bath is
$\agt 1/\omega_0$. In this latter case, the friction force in
Eq.~(\ref{ddot}) is also retarded (cf. \cite{Krivoglaz}).

\subsection{Activation energy of escape}

If the dimensionless noise intensity $D$ is small, for most of the
time the oscillator fluctuates in a small vicinity of one or the other
stable state ${\bf q}_{st}^{(n)}$. Only occasionally there occurs a
large fluctuation which results in a transition to another stable
state. The probability $W_n$ of such fluctuation is exponentially
small, and its dependence on the noise intensity is given by the
activation law, $W_n\propto
\exp(-S_n/D)$ (see \cite{Lindenberg} for a review). In fact, to
logarithmic accuracy $W_n$ is determined by the probability density of
the least improbable realization of the force
$\bbox{\xi}(\tau/\Gamma)$ which results in the corresponding
transition. Therefore the quantity $S_n$ is given by the solution of a
variational problem. This problem is of the form
(cf. \cite{Krivoglaz,Freidlin}):

\begin{eqnarray}
&&W_n=C \exp(-S_n/D),\;
S_n=\min S[{\bf q}(\tau)], \;\label{action}\\ 
&&S[{\bf q}(t)] = \int_{-\infty}^{\infty}
d\tau\, L(\dot{\bf q}, {\bf q}),\;L(\dot{\bf q}, {\bf
q})= {1\over 2}\left[\dot {\bf q}-{\bf K}({\bf
q})\right]^2,\;\nonumber\\ 
&&{\bf q}(t)\rightarrow {\bf
q}_{st}^{(n)}\;{\rm for}\; t\rightarrow -\infty,\quad {\bf
q}(t) \rightarrow {\bf q}_{u} \; {\rm for} \; t\rightarrow \infty,
\nonumber
\end{eqnarray}

\noindent
where the components of the vector ${\bf K}$ are given by the rhs
parts of the equations of motion (\ref{SDE}),

\begin{equation}
K_1({\bf q})= - q_1 +{\partial g\over \partial
q_2},\; K_2({\bf q})= - q_2 - {\partial g\over
\partial q_1}\label{K(q)}
\end{equation}

\noindent
(in what follows we set $n=1,2$ for the stable states of period two
vibrations, and $n=0$ for the stationary state ${\bf q}=0$ where it is
stable).

The solution of the variational problem (\ref{action}) ${\bf q}(t)$
describes the optimal, or most probable escape path (MPEP) from the
$n$th stable state. This path is instanton-like (see \cite{Langer}):
it starts at the stable state ${\bf q}_{st}^{(n)}$ for $t\rightarrow
-\infty$, and for $t\rightarrow \infty$ it approaches the unstable
state ${\bf q}_{u}$ on the boundary of the domain of attraction to
${\bf q}_{st}^{(n)}$ (having reached the boundary, the system makes a
transition to another stable state with a probability $\sim 1/2$).  It
follows from Fig.~1 that, for $-(\zeta^2-1)^{1/2} <
\Omega<(\zeta^2-1)^{1/2}$, 
where the only stable states are period two attractors, escape from
one of them means a transition to the other. For $\Omega > (\zeta^2
-1)^{1/2}$ escape from one of period two attractors means a transition
to the state where period two vibrations are not excited. From this
state, the system makes fluctuational transitions into one or the other
state of period two vibrations.

The
most probable realization of the force is related to the optimal
fluctuational path via Eq.~(\ref{SDE}), $\bbox{\xi}(\tau/\Gamma)=\dot {\bf
q} - {\bf K}$. Optimal fluctuational paths are physically real, they
have been observed in the experiment \cite{prehistory}.

\section{Activation energy of escape in the 
vicinities of the bifurcation points}

The activation energies $S_n$, as defined by (\ref{action}), depend on
two dimensionless parameters of the driven oscillator: the scaled
field strength $\zeta$ and the scaled frequency detuning $\Omega$
(\ref{parameters}). In the general case, $S_n$ may be calculated
numerically (see Sec.~VI). Explicit expressions for $S_n$ can be
obtained in several limiting cases.

In the range of comparatively small nonlinearity (\ref{limitation}),
the oscillator may experience two bifurcations with the varying field
frequency, as seen from Fig.~1.  In the vicinity of a bifurcation
point, one of the motions of the system near the emerging stable
state(s) becomes slow: there arises a ``soft mode''
\cite{Arnold}. Correspondingly, fluctuations near a bifurcation point 
have universal features \cite{Mangel}. 

Near a bifurcation point one can either solve the variational problem
(\ref{action}) explicitly or reduce the system of equations of motion
(\ref{SDE}) to one first order equation. From this equation one can
find not only the exponent, but also the prefactor in the expression
for the escape rate.

The equation for a slow variable $Q$ can be derived from (\ref{SDE})
by appropriately rotating the coordinates: 

\begin{eqnarray}
&&Q=q_1\cos\beta + q_2\sin\beta,
\;P=-q_1\sin\beta + q_2\cos\beta\label{rotation}\\
&& \tan 2\beta = -\Omega_B^{-1}, \;\Omega_B=\mp (\zeta^2
-1)^{1/2}\nonumber
\end{eqnarray}

\noindent
($\Omega_B$ are the bifurcation values of the dimensionless frequency
detuning, see Fig.~1c). For $|\Omega - \Omega_B|\rightarrow 0$ the
dimensionless relaxation time of the variable $Q$ goes to infinity,
whereas that of $P$ is $1/2$, and therefore $P(\tau)$ follows the slow
variable $Q(\tau)$ adiabatically \cite{Mangel}. The fluctuations in
$P$ can be neglected compared to the fluctuations in $Q$. Using the
adiabatic solution for $P$ (i.e., neglecting $\dot P$ and the noise
term in the equation for $\dot P$) we obtain the following equation
for $Q$:

\begin{eqnarray}
&&\dot Q = -dU/dQ + \Xi(\tau), \; \langle\Xi(\tau)\Xi(\tau')\rangle =
D\tilde\delta(\tau-\tau')\label{bif_eq}\\
&&U(Q) = \Omega_B\left[
{1\over 2}(\Omega-\Omega_B)Q^2 -{1\over 4} \zeta^2Q^4\right],\;
\Omega_B|\Omega-\Omega_B|\ll 1.\nonumber
\end{eqnarray}

It is seen from Eq.~(\ref{bif_eq}) that, near the bifurcation point
$\Omega_B = -(\zeta^2-1)^{1/2}$, the system has one stable state for
$\Omega<\Omega_B$ and two symmetrical stable states for $\Omega>
\Omega_B$ (cf. Fig.~1). 
The escape rates from these symmetrical states are the same and are
given by the Kramers expression \cite{Kramers}

\begin{eqnarray}
&&W_{n} = (\sqrt
2/\pi)|\Omega_B|(\Omega-\Omega_B)\exp(-S_n/D),\label{bif_escape}\\
&&S_n=2\left[U(Q^{(n)})-U(Q^{(u)})\right]=
|\Omega_B|(\Omega-\Omega_B)^2/ 2\zeta^2\nonumber
\end{eqnarray}

\noindent 
($Q^{(n)}$ and $Q^{(u)}$ are the values of the coordinate $Q$ in the
stable and unstable states). 

The rate (\ref{bif_escape}) is the rate at which there occur phase
slip transitions between the period two stable states for $\Omega$
close to $\Omega_B= -(\zeta^2-1)^{1/2}$. Eq.~(\ref{bif_escape})
applies for $\exp(-S_n/D)\ll 1$. The activation energy $S_n$ is
quadratic in the distance $|\Omega-\Omega_B|$ to the bifurcation
point. For a given $|\Omega-\Omega_B|$ the dependence of $S_n$ on the
dimensionless field $\zeta$ is given by the factor $(\zeta^2 -
1)^{1/2}/\zeta^2$.

For $0<\Omega-(\zeta^2-1)^{1/2}\ll 1$, Eq.~(\ref{bif_escape})
describes the rate of transitions from the stable state ${\bf
q}_{st}=0$ to each of the period two stable states. The transitions
occur via the appropriate unstable period two state (the one on the
boundary between the domain of attractions to the stable period two
state and the stable state  ${\bf q}=0$).

\section{Activation energies of escape in the small-damping limit}

\subsection{Motion in the absence of dissipation}

Of special interest is the case where the scaled field amplitude
$\zeta$ is large enough so that the dissipation terms $-q_1, -q_2$ in
the rhs of the equations for slow variables (\ref{SDE}) are
comparatively small. In the neglect of these terms and the random
force, Eqs.~(\ref{SDE}) describe conservative
motion of a particle with the coordinate $q_1$ and momentum $q_2$, and
with the Hamiltonian function $g(q_1, q_2)$ (\ref{g}). This particle
moves along closed trajectories shown in Fig.~2 for the auxiliary
variables $X,Y$:

\begin{eqnarray}
&& X = q_1/\zeta^{1/2},\; Y=q_2/\zeta^{1/2},\; \nonumber\\
&&{dX\over d\tilde \tau}
= {\partial G\over \partial Y},\;{dY\over d\tilde \tau}=-
{\partial G\over \partial X},\; \tilde\tau = \zeta\tau.
\label{conservative}\\ 
G(X,Y)
=&&\zeta^{-2}g\left(\zeta^{1/2}X,\zeta^{1/2}Y\right)= {1\over 2}(\mu
-1)X^2 \nonumber\\
&&+{1\over 2}(\mu +1)Y^2 - {1\over 4}(X^2+Y^2)^2,\;\mu = {\Omega\over \zeta}.
\nonumber
\end{eqnarray}

The conservative motion (\ref{conservative}) depends only on one
parameter, $\mu=\Omega/\zeta$, which characterizes the interrelation
between the frequency detuning and the field strength. The shape of
the effective energy $G(X,Y)$ for two different values of $\mu$ is
shown in Fig.~3. The trajectories in Fig.~2 are just the
cross-sections of the surface $G(X,Y)$ by planes $G=$const. The
extrema of the surface $G(X,Y)$ are the fixed points of the system.

For $\mu<-1$ the surface $G(X,Y)$ has one extremum placed at $X=Y=0$,
which corresponds, with dissipation taken into account, to the stable
state with no period two vibrations excited. For $-1<\mu<1$ the
function $G(X,Y)$ has two maxima at $X=0,\,Y=\pm (\mu+1)^{1/2}$ (they
correspond to two period two attractors) and a saddle point at
$X=Y=0$. For $\mu > 1$, in addition to the above maxima the function
$G(X,Y)$ has a minimum at $X=Y=0$ (the maxima and the minimum
correspond to the stable states of the oscillator), and two saddle
points at $X=\pm (\mu -1)^{1/2}, Y=0$. 
The extreme values of $G$ are given by the expressions

\begin{eqnarray}
&&G_{1,2} = {1\over 4}(\mu + 1)^2,\; G_u=0 \;
{\rm for} \;\mu<1;  \label{extremeG}\\
&&G_0=0, \;G_u={1\over 4}(\mu-1)^2\;{\rm for}\;
\mu>1.\nonumber
\end{eqnarray}

\noindent
We note that for $\mu > 1$ the trajectories surrounding the states at
$X=0,\,Y=\pm (\mu+1)^{1/2}$ in Fig.~2 become horse-shoe like for large
enough $G_{1} - G$, and also that for $0\leq G\leq G_u$ there
are coexisting ``internal'' and ``external'' trajectories.

\subsection{Escape rates}

The effect of small dissipation in Eqs.~(\ref{SDE}) is to transform
the closed trajectories in Fig.~2 into small-step spirals which wind
down to the corresponding stable states (cf. Fig.~1). The motion can
be described in a standard way in terms of slow drift over the energy
$g$ towards the stable state (cf. \cite{Lichtenberg}). 

The random force which drives the system away from the stable state
should ``beat'' this drift. It would be expected that the optimal
fluctuational path corresponds to energy diffusion {\it away} from the
stable state. A solution of the variational problem (\ref{action}) for
small dissipation was obtained in \cite{Krivoglaz} for a different
form of the function $g(q_1,q_2)$. Alternatively, one can use the
approach first suggested by Kramers \cite{Kramers} in the analysis of
escape of underdamped thermal equilibrium systems. This approach is
equivalent to deriving from Eqs.~(\ref{SDE}) an equation for $\dot g$
and averaging the dissipative term and the noise intensity in this
equation over the period of vibrations with a given $g$ in the absence
of dissipation and noise.  The dissipative term (the dissipation rate
of $g$) is then given by the expression
$-\overline{[q_1(\partial g/\partial q_1)+q_2(\partial g/\partial q_2)]}$,
with an accuracy to the correction $\propto D$ (here, overline means
averaging over the vibration period).  The diffusion coefficient for
$g$ is given by $D\overline{[\partial g/\partial q_1)^2+(\partial
g/\partial q_2)^2]}$. The resulting first order equation for $g$ can
be solved to give the following expression for the activation energy:

\begin{eqnarray}
&&S_n=2\zeta\,\int_{G_n}^{G_u}dG\,{M(G)\over N(G)},\; 
M(G)=\int\!\!\int_{A(G)}
dXdY,\nonumber\\ 
&&N(G)={1\over 2}\int\!\!\int_{A(G)}dXdY\,\nabla^2 G(X,Y).
\label{underdamped}
\end{eqnarray}
\noindent
Using the explicit form of $G(X,Y)$ one obtains
\begin{equation}
N(G)=\int\!\!\int_{A(G)}dXdY\,\left[\mu -
2(X^2+Y^2)\right],\label{underdampedN}
\end{equation}
\noindent The values of $G_n$ ($n=$1,2) and $G_u$ are the extreme values of $G(X,Y)$ as given
by Eq.~(\ref{extremeG}).  The double integrals in (\ref{underdamped})
are taken over the areas $A(G)$ limited by the trajectories $G(X,Y)=G$
in Fig.~2 which surround the $n$th center (the expressions for $M(G),
N(G)$ were obtained from the expressions for the drift and diffusion
coefficients for $g$ using the Stocks theorem, with account taken of 
Eqs.(\ref{conservative}), as it was done in \cite{Krivoglaz}).

The expressions for $M(G),N(G)$ (\ref{underdamped}) can be further
simplified by changing to polar coordinates
$X=R\cos\varphi,\,Y=R\sin\varphi$. Solving Eq.~(\ref{conservative}) for
$R^2$ in terms of $G, \,\varphi$ and integrating over $R^2$ one then
obtains that, in the problem of
the escape from the period two attractors ($n=1,2$ in
(\ref{underdamped}), i.e., $G\geq G_u$, cf. Fig.~3),

\begin{eqnarray}
&&M(G)=\int d\varphi \,f(G,\varphi),\; \label{simpl1} \\
&&N(G)=-\int d\varphi\,(\mu - 2\cos 2\varphi)f(G,\varphi), \nonumber\\
&& f(G,\varphi)=\left[(\mu -\cos 2\varphi)^2-4G\right]^{1/2}\; (G\geq G_u).\nonumber
\end{eqnarray}

\noindent The limits of the integrals over $\varphi$  are determined
from the condition that $f(G,\varphi)$ be real and $\mu-\cos 2\varphi >
0$.

In the problem of escape from the stable state at ${\bf q}=0$ for
$\mu >1$ ($G\leq G_u$, see Fig.~3) 

\begin{eqnarray}
&&M(G) = \int d\varphi \,R^2(G,\varphi),\;\label{simpl2}\\
&&N(G)=\int
d\varphi \,R^2(G,\varphi)\left[\mu-R^2(G,\varphi)\right]\;(G\leq
G_u)\nonumber\\ 
&&R^2(G,\varphi) = \mu - \cos 2\varphi -
\left[(\mu -\cos 2\varphi)^2-4G\right]^{1/2}.\nonumber
\end{eqnarray}

\noindent
(here, the integral is taken from $0$ to
$\pi$).

\subsection{Explicit limiting expressions for the escape rates}

The expressions for the activation energies are simplified near
bifurcation points (yet not too close to the bifurcation points, so that
the damping of the vibrations with a given $g$ (\ref{conservative}) be
small compared to their frequency). These bifurcation points are $\mu =
\pm 1$. 

As the increasing $\mu$ goes through the value $\mu=-1$, the maximum
of the function $G(X,Y)$ at $X=Y=0$ becomes a saddle point from which
there are split off two maxima of $G$ corresponding to period two attractors
(see Fig.~3). With the further increase in $\mu$, for $\mu=1$ the
point $X=Y=0$ becomes a minimum of $G(X,Y)$ from which there are split
off two saddles of $G$. Escape from the emerging stable states is
determined by small-radius orbits $G(X,Y) = G$. For such orbits
$M(G)/N(G) \approx 1/\mu$ in (\ref{underdamped}), and therefore

\begin{eqnarray} 
S_1=S_2\approx \zeta(\mu + 1)^2/2,\; 0<\mu +1\ll 1, \label{simpl3}\\
S_0 \approx \zeta(\mu -1)^2/2,\; 0<\mu-1 \ll 1.\nonumber
\end{eqnarray}

In the limit of large $\mu$, the activation energy for escape from
period two attractors (\ref{underdamped}) is determined by orbits with
$G_u=(\mu-1)^2/4 \leq G \leq G_{1,2}=(\mu+1)^2/4$. These orbits have a
shape of narrow arcs on the $(X,Y)$-plane. It is seen from
(\ref{simpl1}) that, for such orbits $M(G)/N(G)\approx 1/\mu$, and

\begin{equation}
S_1=S_2\approx 2\zeta, \; \mu \gg 1. \label{large_mu1}
\end{equation}

\noindent
One can also show from (\ref{simpl2}) that 

\begin{equation}
S_0\approx \zeta\mu, \; \mu \gg 1. \label{large_mu0}
\end{equation}

In the general case of arbitrary values of the parameter
$\mu=\Omega/\zeta$ the  activation energies $S_1=S_2, S_0$ could be
found by evaluating the integrals (\ref{underdamped}), (\ref{simpl1}),
(\ref{simpl2}) numerically. The results are shown in Fig.~4. 

It is seen from Fig.~4 that $S_1=S_2$ is quadratic in $(\mu +1)$ for
small $\mu + 1$, and monotonically increases with the increasing
$\mu$.  For large $\mu$, $S_{1,2}$ saturates at $\approx 2\zeta$ . The
activation energy $S_0$ is quadratic in $\mu -1$ for small $\mu -1$,
and then monotonically increases with the increasing $\mu$. We note
that $S_1>S_0$ for $\mu \alt 6.5$. Respectively, for such $\mu$ the
escape rates from the period two attractors are {\it exponentially
smaller} than from the state where the vibrations are not excited. As
a consequence, the stationary population of the period two attractors
$w_1 = w_2$ is exponentially larger than the population $w_0$ of the
steady state:

\begin{equation}
w_1=w_2= (W_0/2W_1)w_0,\; w_1/w_0
\propto\exp[(S_1-S_0)/D].\label{populations}
\end{equation}

For larger frequency detuning ($\Omega =
\mu\zeta$), the steady state becomes more populated, in agreement with
the intuitive physical argument that, as  the field is detuned further
away from the resonance, it is less likely that the period two
vibrations will be excited, for the same field intensity.

\section{Effects of the sixth-order anharmonic term} 

In the experiments \cite{Tan}, because of the structure of the
electrostatic field in the trap for an oscillating electron, the 6th
order anharmonic term in the Hamiltonian of the electron vibrations
could be relatively large (in fact, the 4th order term could be
relatively small), while higher-order terms remain much smaller than
both the 4th and 6th order terms \cite{private}.  The advantage of
suppressing the 4th order term is that the amplitude of the period-two
vibrations becomes larger.  When the 6th order anharmonicity is taken
into account, the equation of motion takes the form

\begin{equation}
{d^2q\over dt^2} + 2\Gamma {dq\over dt}+ \omega_0^2q + \gamma
q^3+\lambda q^5 +qF\cos\omega_F t = \xi(t)\label{an}
\end{equation}

For the model (\ref{an}), the equations of motion in the rotating
frame are again of the form (\ref{SDE}), but now the function
$g(q_1,q_2)$ is given by the expression

\begin{eqnarray}
&&g(q_1,q_2)={1\over 2}\Omega (q_1^2+q_2^2) - {1\over 4} (q_1^2
+ q_2^2)^2{\rm sgn}\,
\gamma \label{ga} \\
&&-{\rho\over 6\zeta} (q_1^2+q_2^2)^3 
 + {1\over 2}\zeta (q_2^2-q_1^2),\nonumber \\
&&\rho={5\lambda F\over 9 \gamma^2}. \nonumber
\end{eqnarray}
\noindent
We have neglected the renormalization ($\propto \gamma^2$) of
the nonlinearity parameter parameter $\lambda$ (this renormalization
is substantial when $\gamma$ is not small, in which case the role of the
6th order anharmonicity is insignificant, or the whole approximation
of small-amplitude vibrations does not apply). The dimensionless
parameter $\rho$ characterizes the ``strength'' of the 6th order
nonlinearity. 

For $\Omega$ very close to the bifurcation values $\mp
(\zeta^2-1)^{1/2}$, escape from the small-amplitude stable state(s) is
determined by the motion with small $q_1^2+q_2^2$. Clearly,
this motion is determined by the {\it lowest-order} anharmonicity, and
therefore neither this motion nor the bifurcation values of the
parameters are affected by the higher-order nonlinear terms.

In what follows we will investigate the effect of the term $\propto
\lambda$ in (\ref{an}) on the escape rates from the period two
attractors, and also from the zero amplitude stable state for $\Omega
> (\zeta^2-1)^{1/2}$, far from the bifurcation points. We will analyze
the case of weak damping, which is of utmost interest for the
experiment \cite{private}.

In the neglect of dissipation and fluctuations, the motion of the
oscillator (\ref{an}) in the rotating frame can be described by
Eqs.~(\ref{conservative}), but when the 6th order nonlinearity is
taken into account, the effective energy $G(X,Y)$ is of the form

\begin{eqnarray}
&& G(X,Y)= {1\over 2}(\mu -1)X^2 +{1\over 2}(\mu +1)Y^2, \label{Ga} \\
&& - {1\over 4}(X^2+Y^2)^2 -{1\over 6}\rho (X^2+Y^2)^3 ,\;\quad
\mu = {\Omega\over \zeta}. \nonumber
\end{eqnarray}
\noindent

For $\rho >0$ (i.e., for $\gamma\lambda > 0$ in Eq.~(\ref{an})), the
phase portrait of the conservative motion (\ref{conservative}),
(\ref{Ga}) remains qualitatively the same as that shown in Figs.~2a,b
for $\rho=0$, as does also the topological structure of the surface
$G(X,Y)$ in Fig.~3. The centers which correspond to the period two
attractors lie on the $Y$-axis, $X_{1,2}=0$, they correspond to the
maxima of the function $G(X,Y)$, whereas the hyperbolic points which
correspond to the unstable period two states lie on the $X$-axis,
$Y_{u1,2}=0$; they correspond to the saddles of $G(X,Y)$.

The general expression for the activation energy of escape $S_n$ is
given by Eq.~(\ref{underdamped}), but now the functions $M(G)$ and $N(G)$
have to be calculated with account taken of the 6th order terms in
$G(X,Y)$ (\ref{Ga}). In particular, 

\begin{eqnarray}
N(G)=&&\int\!\!\int_{A(G)}dXdY\,\left[\mu -
2(X^2+Y^2)\right.\nonumber \\
&&-\left.3\rho (X^2+Y^2)^2\right]   \label{underdamped1} 
\end{eqnarray}

The functions $M(G), N(G)$ can be written as single integrals
over the polar angle (cf. Sec.~IV~B). In the problem of escape from
period two attractors, we obtain that, similar to (\ref{simpl1}),

\begin{eqnarray}
&&M(G)={1\over 2}\int%_{\varphi_1}^{\varphi_2}
d\varphi\,\left[R_{+}^2(G,\varphi)- R_{-}^2(G,\varphi)\right]
\label{simpl1a} \\
&&N(G)={1\over
2}\int%_{\varphi_1}^{\varphi_2}
d\varphi\,\left[Z_{+}(G,\varphi)-Z_{-}(G,\varphi)\right],
\nonumber \\ 
&&Z_{\pm}(G,\varphi)=\mu
R_{\pm}^2(G,\varphi)-R_{\pm}^4(G,\varphi)-
\rho R_{\pm}^6(G,\varphi) \nonumber 
\end{eqnarray}
\noindent
Here, $R_{\pm}(G,\varphi)$ are the radii of the trajectories
$G(X,Y)=G$ which surround the centers corresponding to the period two
attractors. They are given by the real roots of the equation
\begin{eqnarray}
&&{1\over 6}\rho R_{\pm}^6 +{1\over 4}R_{\pm}^4-{1\over 2} (\mu-\cos
2\varphi)R_{\pm}^2 +G=0, \label{eqR}\\ 
&&G_{1,2} > G> 0 \;{\rm for} \;
\mu < 1, \;G_{1,2}> G > G_u \; {\rm for} \; \mu > 1.\nonumber
\end{eqnarray}

\noindent
($G_1=G_2$ are the values of $G$ in the period two attractors, see
Eq.~(\ref{extreme_G})).The limits of the integrals over $\varphi$ in
Eq.~(\ref{simpl1a}) are determined from the condition that
Eq.~(\ref{eqR}) has two real roots, $R_- < R_+$.

For the case of escape from the attractor with zero vibration
amplitude ({\bf q}=0)  the functions $M(G)$
and $N(G)$ have the form

\begin{eqnarray}
&&M(G)={1\over 2}\int_{0}^{2\pi}d\varphi \, R_{-}^2(G,\varphi), 
\label{simpl2a}\\
&&N(G)={1\over 2}\int_{0}^{2\pi}d\varphi \, Z_{-}(G,\varphi),\quad
\mu > 1,\; G_u> G> 0 \nonumber 
\end{eqnarray}
\noindent
where the function $Z_{-}(G,\varphi)$ is defined in
Eq.~(\ref{simpl1a}) (here, $G_u$ is the value of $G$ in the saddle
points, see Eq.~(\ref{extreme_G}); in the range $0<G<G_u$
Eq.~(\ref{eqR}) has only one real root $R_-$, which determines the
function $Z_-$).

\subsection{Activation energies of escape near bifurcation points}

For underdamped systems where the sixth order anharmonicity, the
analysis of the activation energies of escape for parameter values
close to the bifurcation points is similar to that in Sec.~IV~C.  In
the range $0< \mu+1 \ll 1$ the centers ($X_{1,2},Y_{1,2})$) which
correspond to the period two attractors are close to the saddle point
at the origin. In this case $R_{\pm}\ll 1$ in Eqs.~(\ref{eqR}), and
the terms with $X^2+Y^2$ in Eq.~(\ref{underdamped1}) can be
neglected. These terms can be also neglected in the problem of escape
from the stable state $X=Y=0$ for $0<\mu + 1\ll 1$, as in this case
the hyperbolic points $(X_{u1,2},Y_{u1,2})$ are close to the stable
state. Therefore, in both cases, $M(G)/N(G) \approx 1/\mu$
(cf. ~Sec.~IV~C), and it follows from Eq.~(\ref{underdamped}) that the
activation energies of escape

\begin{eqnarray}
&&S_{1}=S_{2} \approx 2\zeta G_1, \quad 0< \mu+1 \ll 1,
\label{activ} \\ 
&&S_0\approx 2\zeta G_u\quad 0< \mu-1 \ll 1
\nonumber
\end{eqnarray} 

\noindent
Here, we have taken into account that $G(0,0)=0$. The explicit
expressions for the effective energies $G_1=G_2$ (the maxima of
$G(X,Y)$) and $G_u\equiv G_{u1}=G_{u2}$ (the saddles of $G(X,Y)$) are
of the form

\begin{eqnarray}
&& G_{1,2}={[-(1+6\rho (\mu+1)]+[1+4\rho(\mu+1)]^{3/2}
\over 24\rho^2}, \label{extreme_G}\\
&&G_{u}={[-(1+6\rho
(\mu-1)]+[1+4\rho(\mu-1)]^{3/2}
\over 24\rho^2}.\nonumber
\end{eqnarray}

For weak sixth order nonlinearity $\rho \ll 1$, or just very close to
the bifurcation points, so that $\rho (\mu^2- 1)\ll 1$ (but the
effects of dissipation are still small), the expressions
(\ref{activ}), (\ref{extreme_G}) go over into the
asymptotic expressions for the activation energies (\ref{simpl3}),
$S_n \propto (\mu^2-1)^2$.

\subsubsection{Strong sixth order nonlinearity}

The activation energies (\ref{activ}) as functions of the distance
$\mu^2-1$ to the bifurcation points display an interesting behavior
for large sixth-order nonlinearity, $\rho \gg 1$, in the range where
$\mu^2-1\ll 1$ but $\rho (\mu^2-1) \gg 1$. It follows from
(\ref{activ}), (\ref{extreme_G}) that in this range

\begin{eqnarray}
&&S_{1,2} \approx {2\over 3}\zeta(\mu + 1)^{3/2}\rho^{-1/2},\;
\mu+1\ll 1,\quad  \rho (\mu+1) \gg 1, \label{bifur6}\\
&& S_0\approx {2\over 3}\zeta(\mu -1)^{3/2}\rho^{-1/2},\;
\mu - 1\ll 1, \quad  \rho (\mu-1) \gg 1. \nonumber
\end{eqnarray}
\noindent
The dependence on the distance to the bifurcation point $\mu^2-1$ as
given by Eq.~(\ref{bifur6}) is described by the power law with the
exponent $3/2$. In contrast, for weak sixth order nonlinearity
(\ref{simpl3}) the exponent is equal to 2. Eqs.~(\ref{activ}),
(\ref{extreme_G}) describe both limiting behaviors and the crossover
from one of them to the other.

\subsection{Strong sixth order nonlinearity:  general case}

For strong sixth order nonlinearity, $\rho \gg 1$ and $\rho(\mu^2-1)
\ll 1$, it is convenient to rescale the dynamical variables,

\begin{eqnarray}
&&\tilde X= \rho^{1/4}X,\;\tilde Y= \rho^{1/4}Y, \;\tilde G=
\rho^{1/2}G \label{rescale1}\\
&&\tilde G = {1\over 2}(\mu -1)\tilde X^2 +{1\over 2}(\mu +1)\tilde Y^2
 -{1\over 6} (\tilde X^2+\tilde Y^2)^3, \nonumber
\end{eqnarray}

\noindent
and the noise intensity $D$ (\ref{correlator}),

\begin{equation}
\tilde D= \rho^{1/2}D={(5\lambda F)^{1/2}\over 8\omega_F^3\Gamma^2}
\phi(\omega_F/2). \label{tildeD}
\end{equation}

The maximal and saddle values of $\tilde G$, $\tilde G_{1,2}$ and
$\tilde G_u$, respectively, are given by the simple expressions

\begin{eqnarray}
&&\qquad G_{1,2}={1\over 3}(\mu+1)^{3/2};\label{Gext} \\
&&G_u=0\; {\rm for}\; -1<\mu<1, \quad G_u={1\over 3}(\mu-1)^{3/2}
\; {\rm for}\; \mu>1, \nonumber
\end{eqnarray}

\noindent
whereas the expressions for the trajectories in polar coordinates
$(\tilde R\equiv \rho^{1/4}R,\varphi)$ are of the form

\begin{eqnarray}
&&\tilde R_{\pm}^{2}(\tilde G,\varphi)=2(\mu-\cos 2\varphi)^{1/2}
\cos\left[{\theta(\tilde G,\varphi)\mp \pi\over 3}\right], \label{eqR6} \\
&& \theta(\tilde G,\varphi)= \arccos\left[3\tilde G/(\mu-\cos
2\varphi)^{3/2}\right] \nonumber
\end{eqnarray}

With account taken of (\ref{eqR6}), the expressions (\ref{simpl1a})
for the functions $\tilde M =\rho^{-1/2}M,\tilde N = \rho^{-1/2}N$
take the form

\begin{eqnarray}
\tilde M(\tilde G)=&&\int d \varphi [3(\mu-\cos 2\varphi)]^{1/2}\sin
[\theta(\tilde G,\varphi)/3] ,
\label{M6} \\
\tilde N(\tilde G)=&&-\int d\varphi (2\mu -3\cos 2\varphi)\left[3(\mu-\cos
2\varphi)\right]^{1/2} \nonumber \\ 
&& \times \sin [\theta(\tilde G,\varphi)/3] ,
\nonumber 
\end{eqnarray}

\noindent
(the above expressions refer to the trajectories which surround the
period two attractors). For the trajectories that surround the
zero-amplitude state $\tilde X=\tilde Y =0$ we obtain from
(\ref{simpl2a}), (\ref{eqR6})

\begin{eqnarray}
\tilde M(\tilde G)=&&\int_0^{2\pi} d \varphi\, [\mu-\cos
2\varphi]^{1/2}\cos
\left[{\theta(\tilde G,\varphi)+\pi\over 3}\right] ,
\label{M06} \\
\tilde N(\tilde G)=&&-\int_0^{2\pi} d\varphi\,
(2\mu-3\cos2\varphi)[\mu-\cos 2\varphi]^{1/2} \nonumber \\
&&\times\cos \left[{\theta(G,\varphi)+\pi\over 3}\right]+
6\pi \tilde G.
\nonumber
\end{eqnarray}

The expression for the escape activation energies $\tilde
S_n=\rho^{1/2}S_n$ has the same form as Eq.~(\ref{underdamped}) for
$S_n$, with $G,M,N$ in (\ref{underdamped}) replaced by $\tilde
G,\tilde M, \tilde N$, respectively.  The escape rate in the variables
with tilde has the same form as in Eq.~(\ref{action}),

\begin{equation}
W_n=C \exp(-\tilde S_n/\tilde D). \label{rescale2}
\end{equation}

\noindent
It follows from Eqs.~(\ref{ga}), (\ref{rescale1}), (\ref{rescale2})
that, in the limit of large sixth-order nonlinearity, the fourth-order
nonlinearity parameter $\gamma$ drops out of the expressions for the
activation energies $\tilde S_n$, the reduced noise intensity $\tilde
D$, and the escape rates. This is in agreement with
Eq.~(\ref{bifur6}), which shows explicitly that $S_n\propto
\rho^{-1/2}$ for  large $\rho$, and therefore $\tilde S_n=\rho^{1/2}S_n$ 
is independent of $\gamma$.

For large frequency detuning, $\mu \gg 1$, it follows from
Eqs.~(\ref{underdamped}), (\ref{M6}), (\ref{M06}) that the activation
energies $\tilde S_{1,2}, \tilde S_0$ are of the form

\begin{equation}
\tilde S_{1,2}\approx \zeta \mu^{-1/2}, \quad  
\tilde S_0 \approx \zeta \mu^{1/2},\;
\quad (\mu\gg 1). \label{asS_n}
\end{equation}

It is clear from the asymptotic expressions (\ref{bifur6}),
(\ref{asS_n}) that the activation energy of escape from the period two
attractors $\tilde S_{1,2}$ is a {\it nonmonotonic} function of $\mu$,
i.e. of the frequency detuning $\omega_F-2\omega_0$. The decrease of
$\tilde S_{1,2}$ for large $\mu$ can be understood as follows. As we
mentioned before, the effective reciprocal ``temperature'' of the
distribution of the system over the energy $G$, which is given by the
$M/N=\tilde M/\tilde N$, is determined by the ratio of the rates of
the drift and diffusion of the oscillator over $G$ (for a Brownian
particle this ratio is indeed equal to $1/kT$). The drift coefficient
is linear in the characteristic velocity $\dot X, \dot Y$ of the
oscillator in rotating frame, whereas the diffusion is quadratic in
this velocity. Therefore when the velocity is large, the ratio $M/N$
becomes small. This happens for large $\mu$. Here, the characteristic
time scale for the motion with a given $G$ is set by the reciprocal
frequency detuning $1/|\omega_F-2\omega_0|$, and the ratio $M/N\propto
1/|\omega_F-2\omega_0|\propto 1/\mu$. On the other hand, the energy
interval $G_{1}- G_u$ for large $\mu$ is increasing with $\mu$
sublinearly, if the sixth order nonlinearity is dominating (and
linearly, if the fourth order nonlinearity is dominating). Therefore,
for large $\rho$ and $\mu$ the activation energy decreases with the
increasing $\mu$.

The position of the maximum of $\tilde S_{1,2}$ and the overall
dependence of $\tilde S_n$ on $\mu$ can be obtained by numerical
integration of Eqs.~(\ref{underdamped}), (\ref{M6}), (\ref{M06}). The
results are shown in Fig.~5.

\acknowledgments
  
The research of C.M. Maloney and M. Silverstein was supported through
the REU program at Michigan State University. M.I. Dykman acknowledges
support from the NSF Grant no. PHY-9722057.

\appendix
\section{Weak sixth order nonlinearity}
When $\rho \zeta \ll 1$ the zeroth-order in $\rho \zeta$ values of the
functions $M(G)$ and $N(G)$ in the Eqs.~(\ref{simpl1a}) and
(\ref{simpl2a}) are determined by the Eqs.(\ref{simpl1}) and
(\ref{simpl2}), respectively.  In the case of period two attractors to
the first order in $\rho \zeta$ the functions $M$, $N$ have the form

\begin{eqnarray}
&&M(G)\approx M^{(0)}(G)+M^{(1)}(G),N(G)\approx
N^{(0)}(G)+N^{(1)}(G). \nonumber \\ &&M^{(1)}(G)=-{4\rho \zeta\over 3}
\int_{\varphi_1}^{\varphi_2} d\varphi\,
{1\over f(G,\varphi)}(\mu-\cos2\varphi) \label{M1} \\
&&\times [(\mu -\cos 2\varphi)^2-3\tilde G], \nonumber \\
&&N^{(1)}(G)=N^{(1)}_{1}(G)+N^{(1)}_{2}(G) , \label{N1} \\
&&N^{(1)}_{1}(G)=-4\rho \zeta \int_{\varphi_1}^{\varphi_2} d\varphi\,
f(G,\varphi)[(\mu-\cos2\varphi)^2-G], \nonumber \\
&&N^{(1)}_{2}(G)={\rho \zeta\over 6}\int_{\varphi_1}^{\varphi_2}
d\varphi\,\left[2\left((R_{+}^{(0)}(G,\varphi))^7+
(R_{-}^{(0)}(G,\varphi))^7\right)\right. \nonumber
\\ &&\left.-\mu
\left((R_{+}^{(0)}(G,\varphi))^5+
(R_{-}^{(0)}(G,\varphi))^5\right)\right]/f(G,\varphi)
\nonumber
\end{eqnarray}
\noindent

Here $R_{\pm}^{(0)}$ are the radii of the orbits with a given $G$
evaluated for $\rho = 0$. The angles $\varphi_{1,2}$ in
(\ref{M1}),(\ref{N1}) are evaluated in the zeroth order in $\rho
\zeta$

\begin{equation}
\varphi_1={1\over 2}\arccos (\mu-2G^{1/2}), \,\varphi_2=\pi-\varphi_1,
\end{equation}

%\begin{eqnarray}
%\Phi_{\omega}[\xi_1(t+t')+i\xi_2(t+t'),&&\xi_1(t)-i\xi_2(t)] \nonumber\\
%&&={3|\gamma|\over 4\omega_F^3\Gamma}\Phi\left(\omega- {1\over
%2}\omega_F\right)\label{corr1}\\
%\Phi_{\omega}[\xi_1(t+t')+i\xi_2(t+t'),&&\xi_1(t)+i\xi_2(t)]\nonumber\\
%&& =
%{3|\gamma|\over 4\omega_F^3\Gamma}e^{-i\omega_Ft}\Phi\left(\omega- {1\over
%2}\omega_F\right)\nonumber
%\end{eqnarray}

\newpage

\centerline{\bf Figures}

\begin{center}
\epsfxsize=3.0in                %so many inches wide
\leavevmode\epsfbox{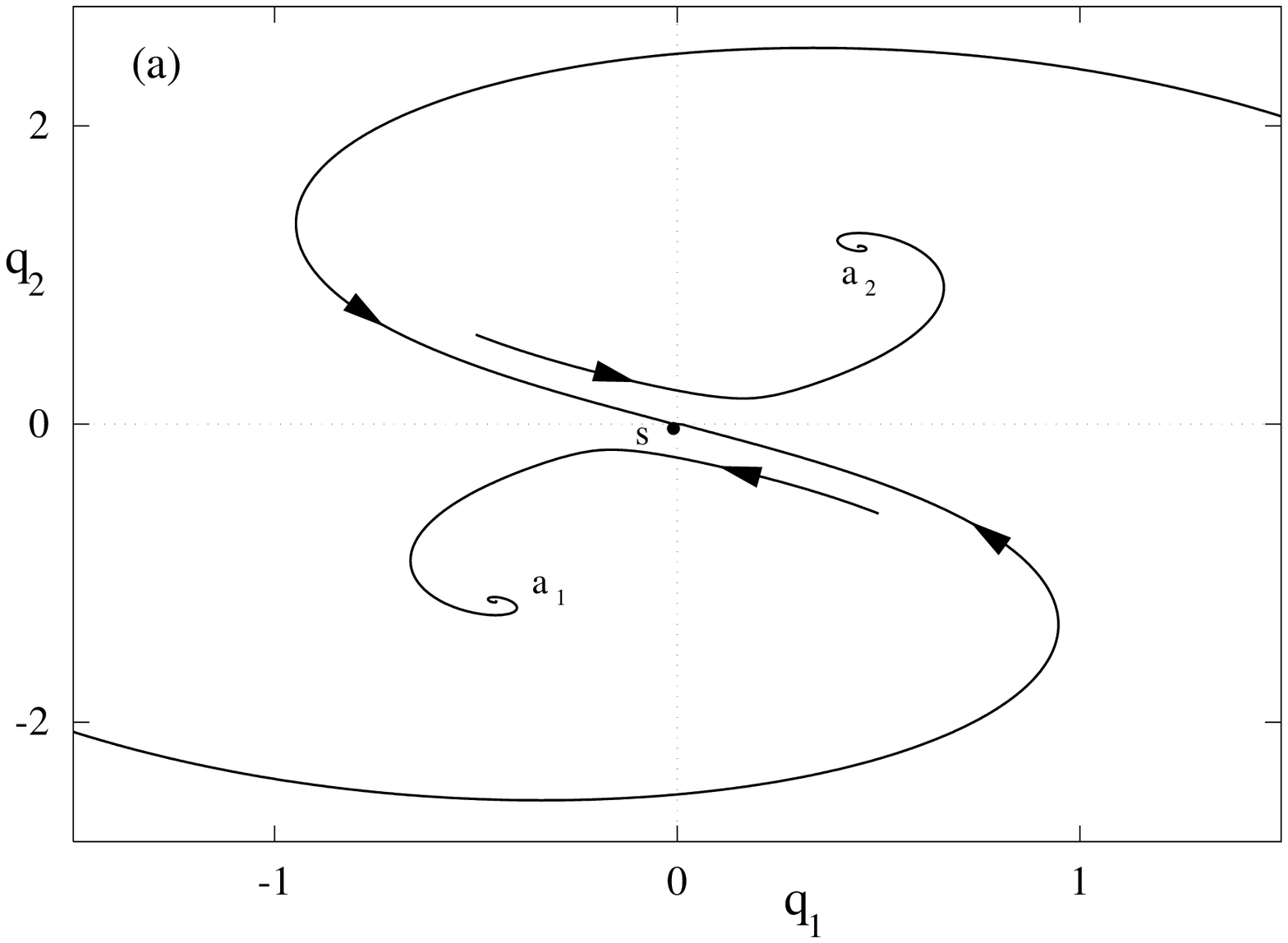}
\end{center}
\begin{center}
\epsfxsize=3.0in                %so many inches wide
\leavevmode\epsfbox{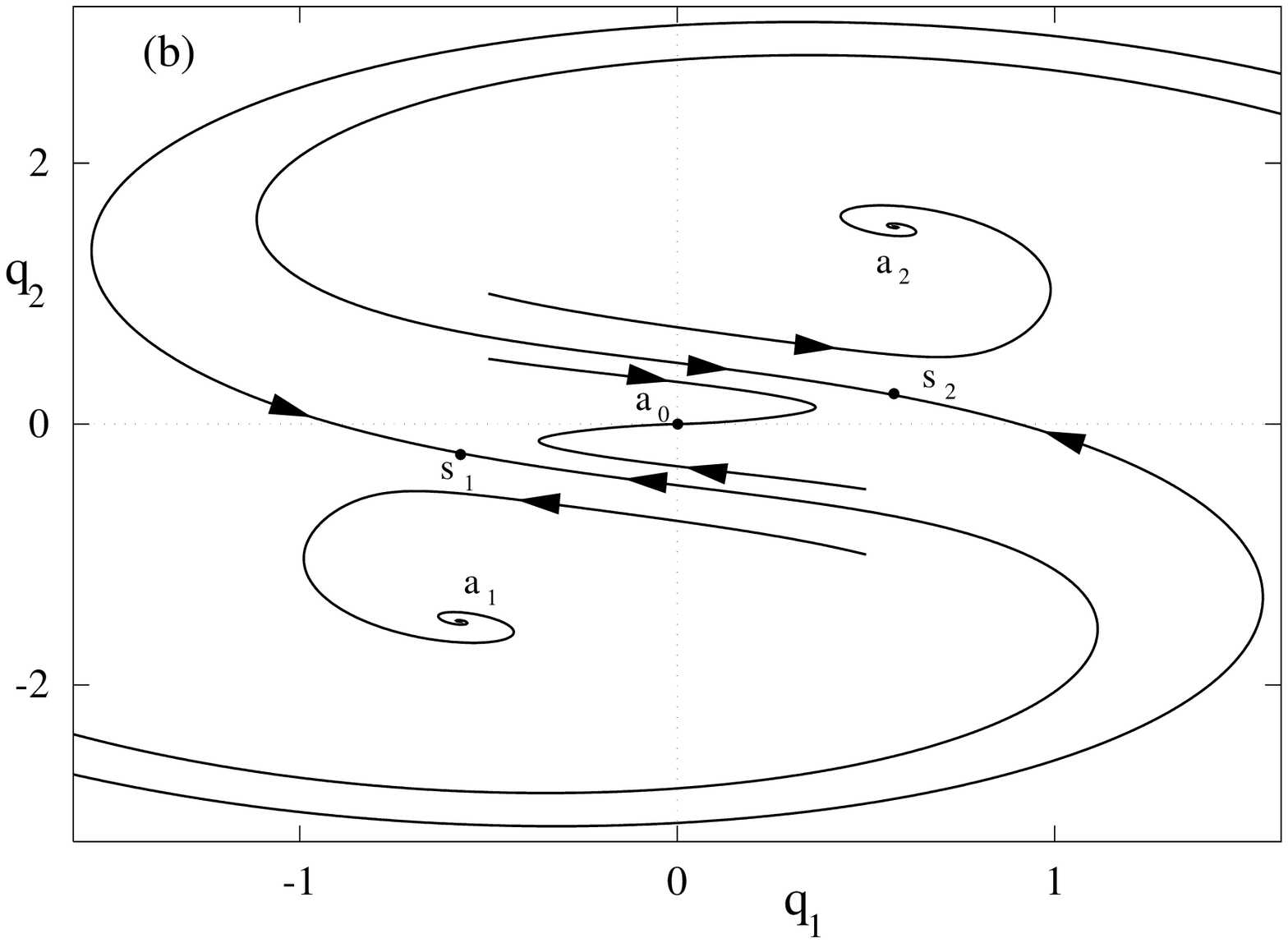}
\end{center}
\begin{center}
\epsfxsize=3.0in                %so many inches wide
\leavevmode\epsfbox{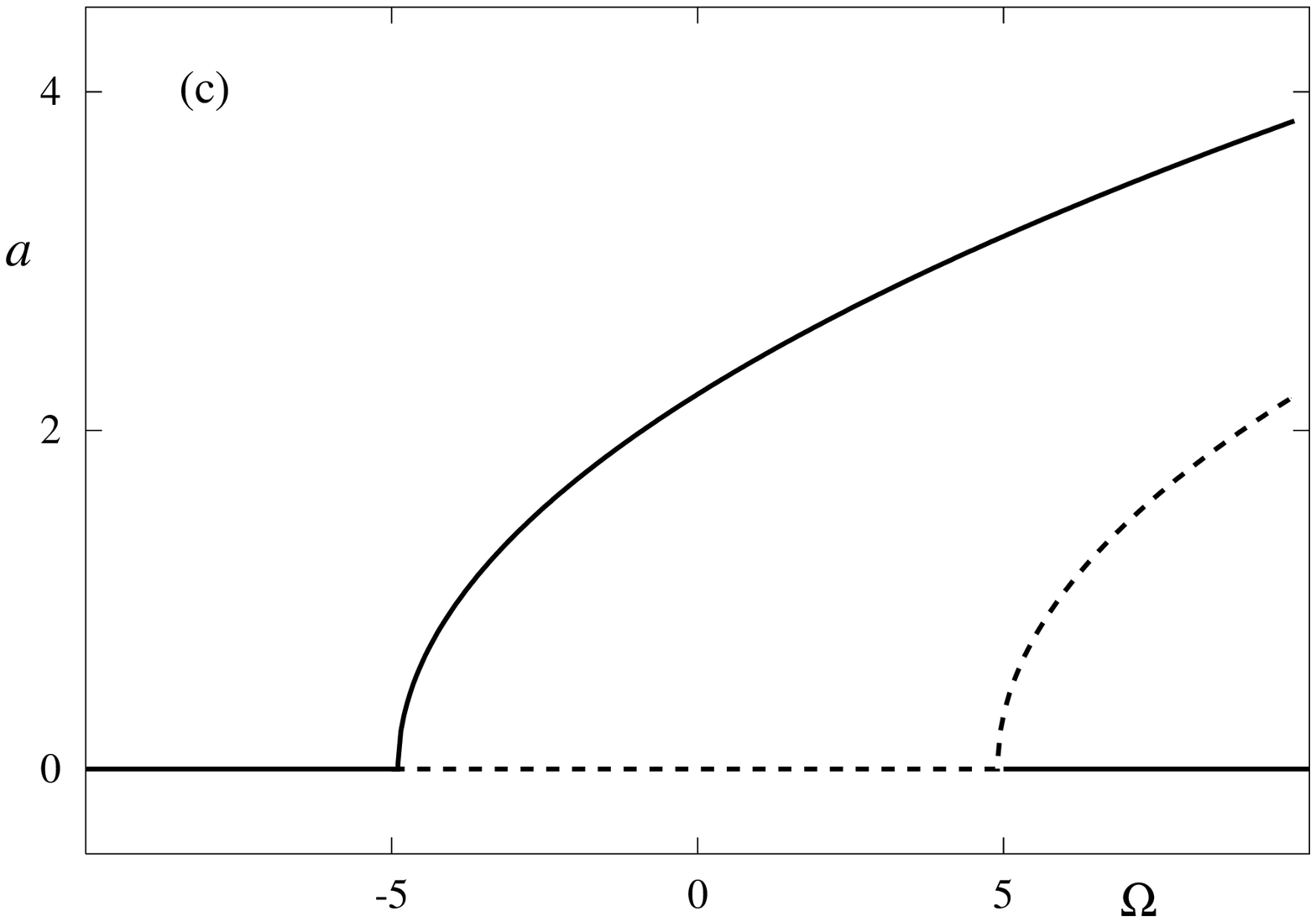}
\end{center}

\begin{figure}
\caption{Trajectories and separatrices of the oscillator in the absence of
noise in slow variables $q_1, q_2$ for (a) $\zeta = 1.5, \,\Omega =
0.5$ where the stable states of the oscillator are period two
attractors, and (b) for $\zeta = 1.5, \,\Omega = 1.5$ where the steady
state ${\bf q}=0$ is also stable. The positions of the stable states
and the saddle points are denoted by the letters $a_i$ and $s$,
respectively. (c) The dependence of the dimensionless amplitude of the
stable (solid line) and unstable (dashed line) period two vibrations
on $\Omega$ for $\zeta = 5.0$.}
\end{figure}

\begin{center}
\epsfxsize=2.8in                %so many inches wide
\leavevmode\epsfbox{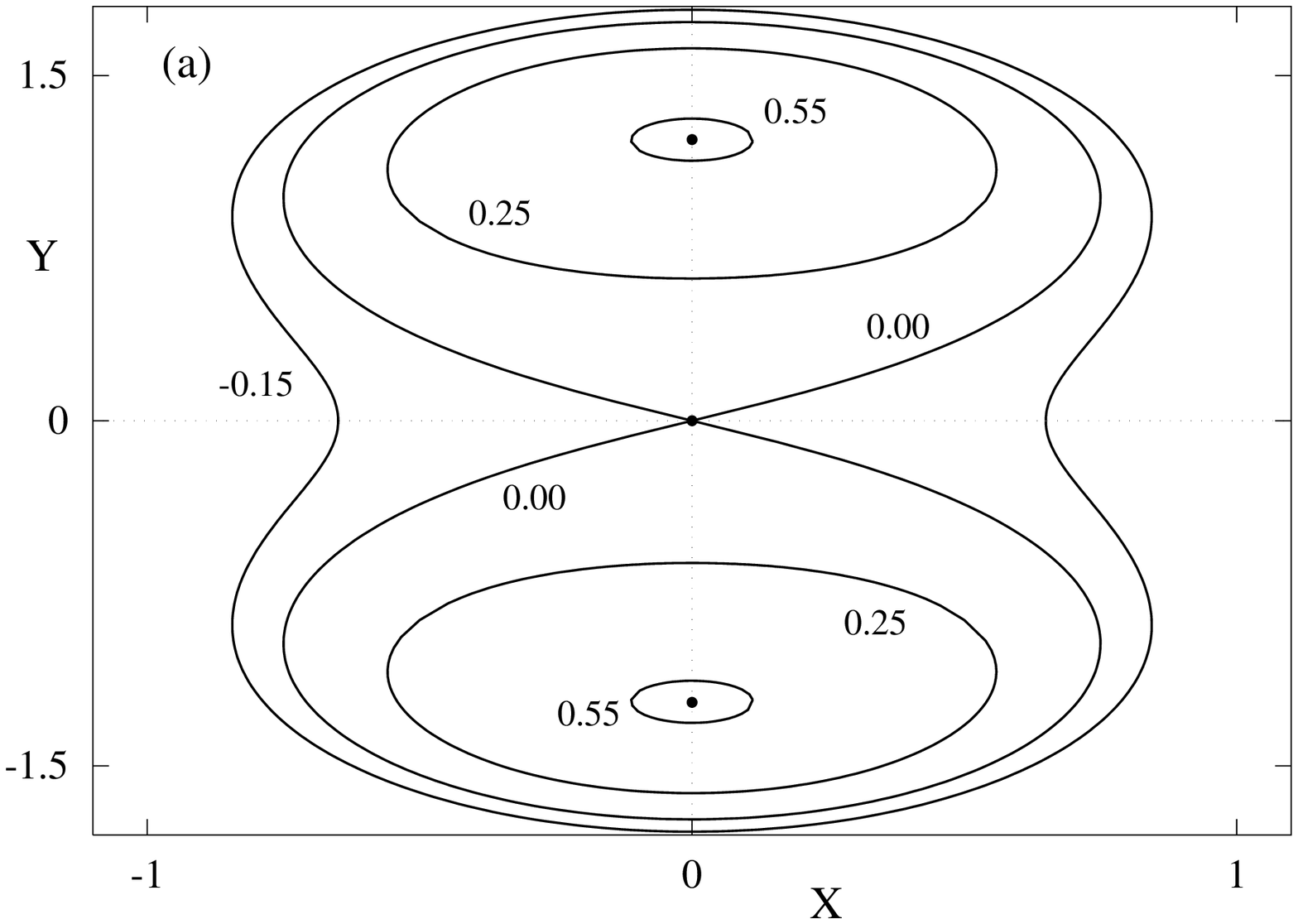}
\end{center}
\begin{center}
\hspace{-.44in}
\epsfxsize=3.2in                %so many inches wide
\leavevmode\epsfbox{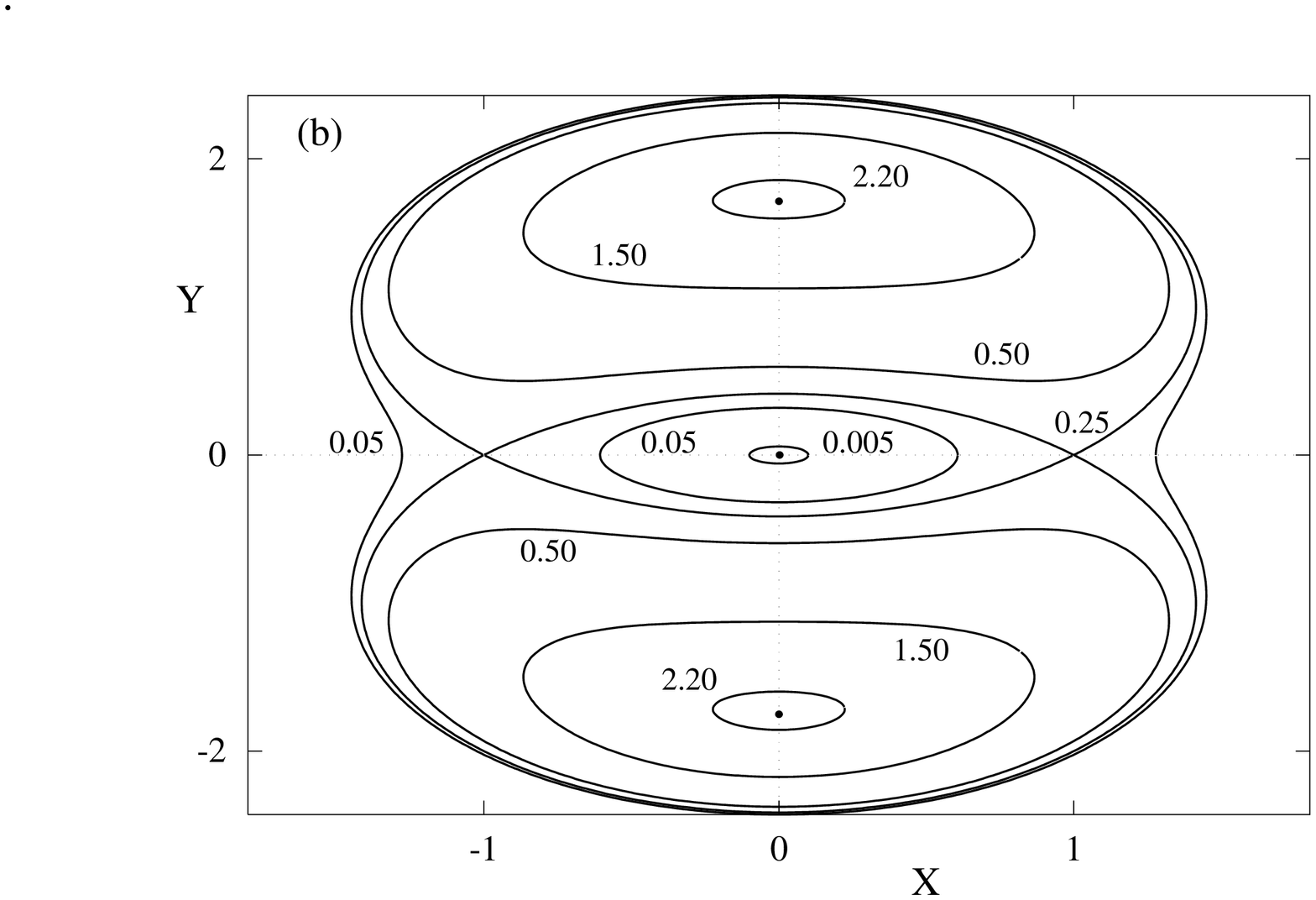}
\end{center}

\begin{figure}
\caption{Trajectories of the conservative 
motion (16) for (a) $\mu = 0.5$ and (b) $\mu = 2.0$. The values of the
Hamiltonian function $G(X,Y)$ are shown near the trajectories. The
dots show the positions of the centers and the hyperbolic points.}
\end{figure}

\vspace{-1.2in}
\begin{center}
\epsfxsize=3.2in                %so many inches wide
\leavevmode\epsfbox{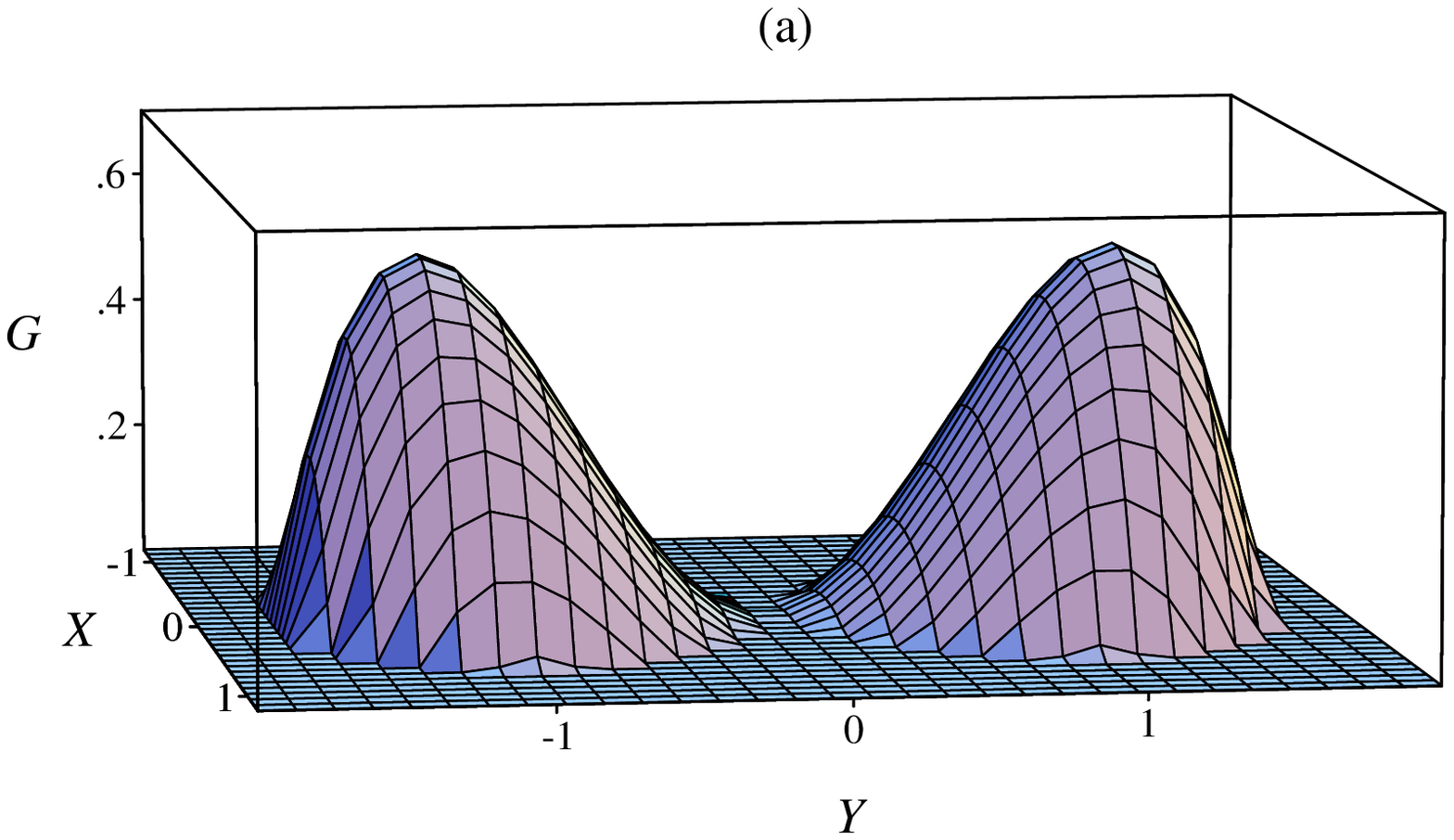}
\end{center}
\vspace{-2.4in}
\begin{center}
\epsfxsize=3.2in                %so many inches wide
\leavevmode\epsfbox{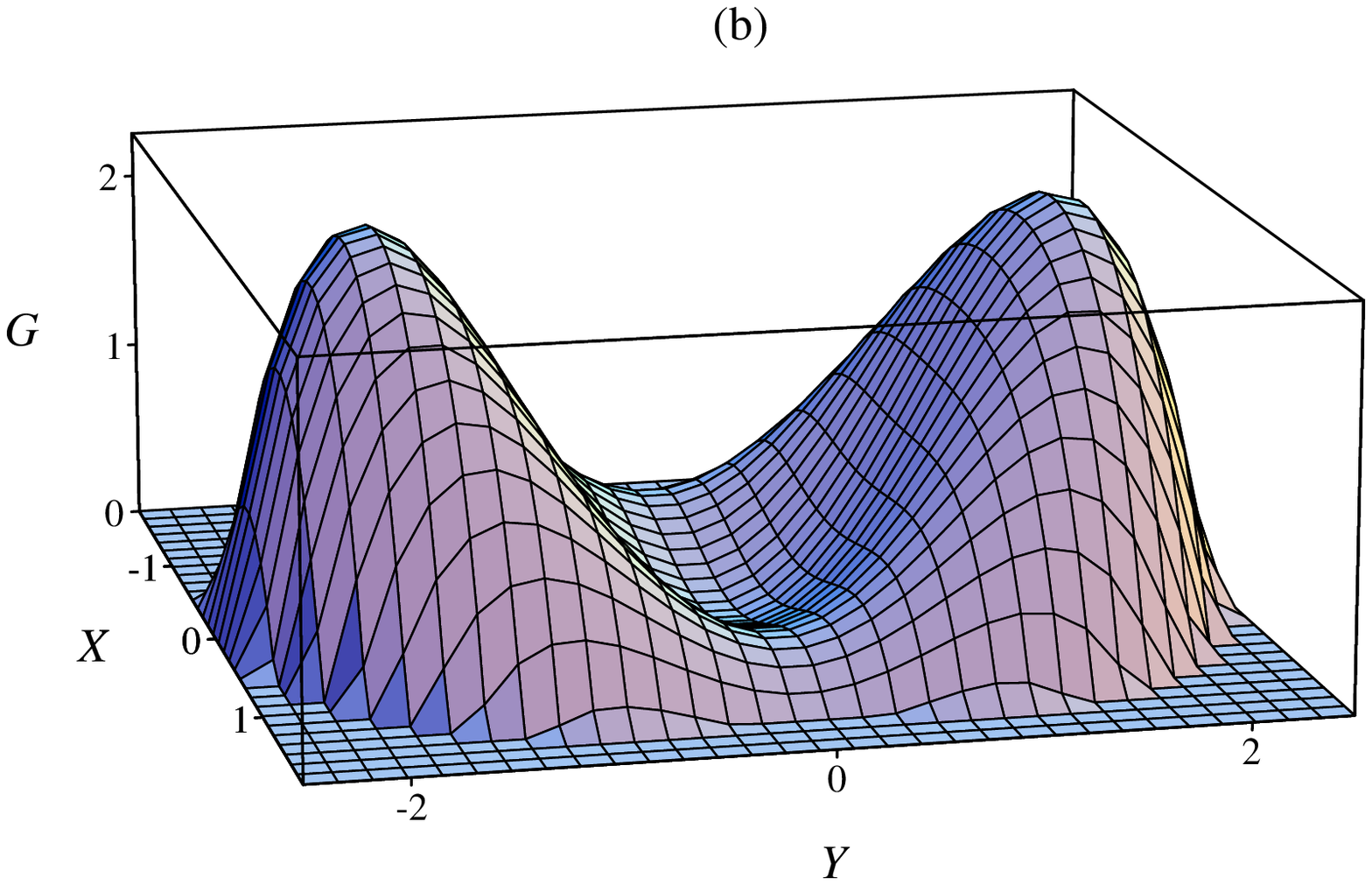}
\end{center}
\vspace{-1.2in}

\begin{figure}
\caption{The Hamiltonian function $G(X,Y)$ (16) (a) for $\mu = 0.5$ 
where the function $G$ has two maxima (which correspond to period two
attractors, with account taken of dissipation) and a saddle point at
$X=Y=0$, and (b) for $\mu=2.0$ where $G$ has two maxima, a minimum at
$X=Y=0$, and two saddles which correspond to unstable period two
vibrations, with account taken of dissipation.}
\end{figure}

\begin{center}
\epsfxsize=3.0in                %so many inches wide
\leavevmode\epsfbox{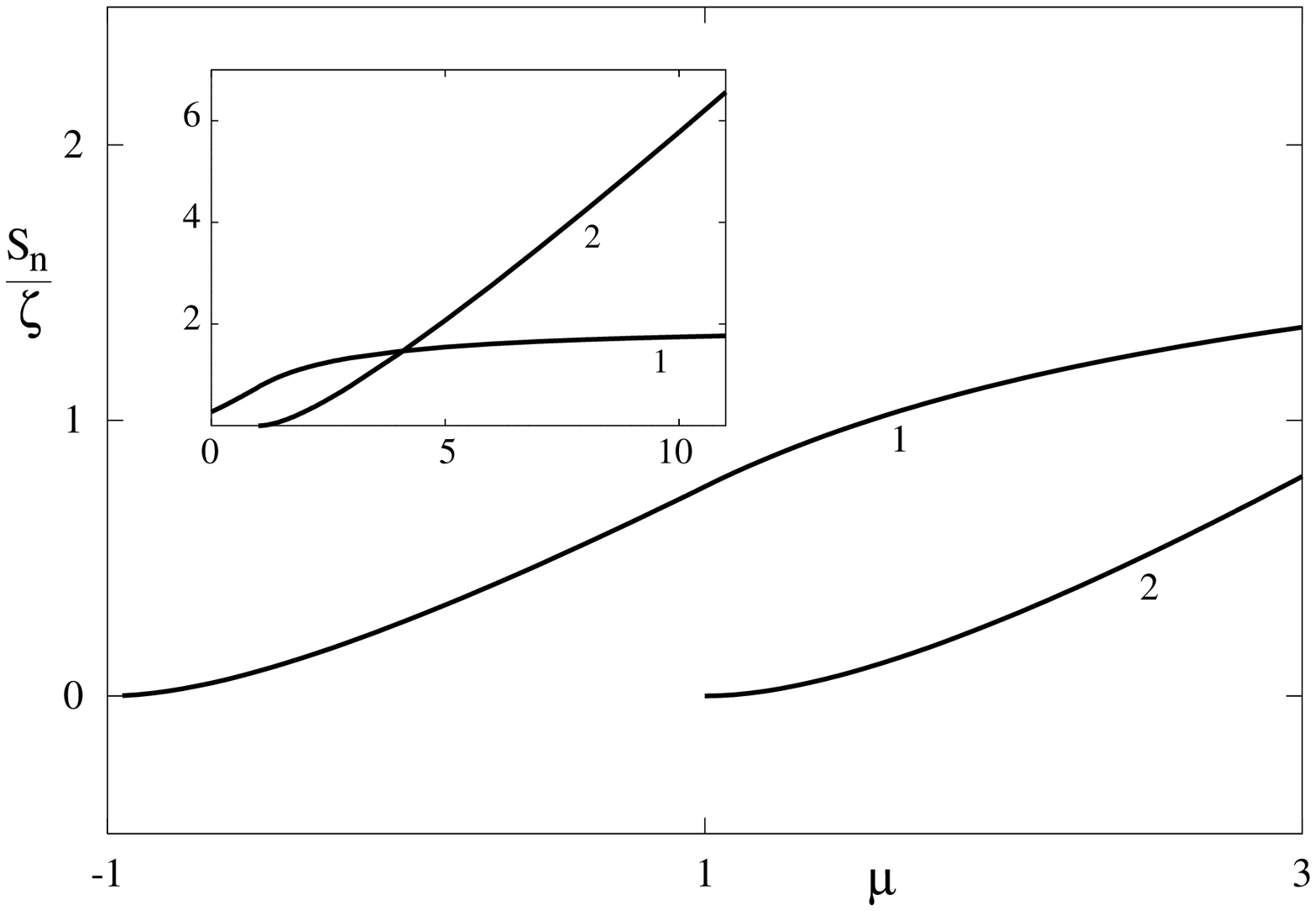}
\end{center}

\begin{figure}
\caption{The dependence of the escape activation energy $S_1=S_2$ 
(lines 1) and $S_0$ (lines 2) on the scaled frequency detuning $\mu=
\Omega/\zeta \equiv 2\omega_F[(\omega_F/2)-\omega_0]/F$ in the
limit of comparatively large fields or small damping, $\zeta\gg 1$. }
\end{figure}

\begin{center} \epsfxsize=3.0in %so many inches wide 
\leavevmode\epsfbox{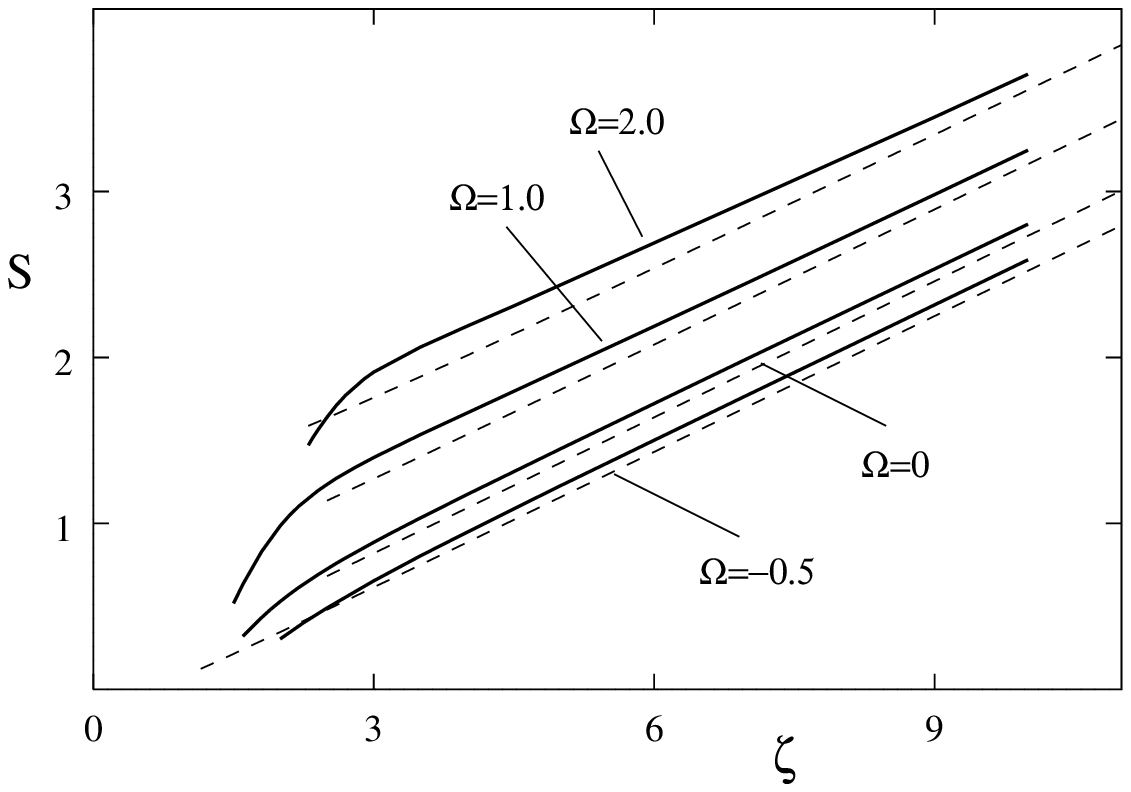}
\end{center} 

\begin{figure}
\caption{The dependence of the activation energies 
of the phase slip transitions between period two attractors on the
scaled field amplitude for two different values of the frequency
detuning $\Omega$, as obtained by solving the variational problem (11)
(solid lines).  The dashed lines show the low-damping (large $\zeta$)
asymptotes (cf. Fig.~4)}
\end{figure}

\begin{center}
\epsfxsize=3.0in                %so many inches wide
\leavevmode\epsfbox{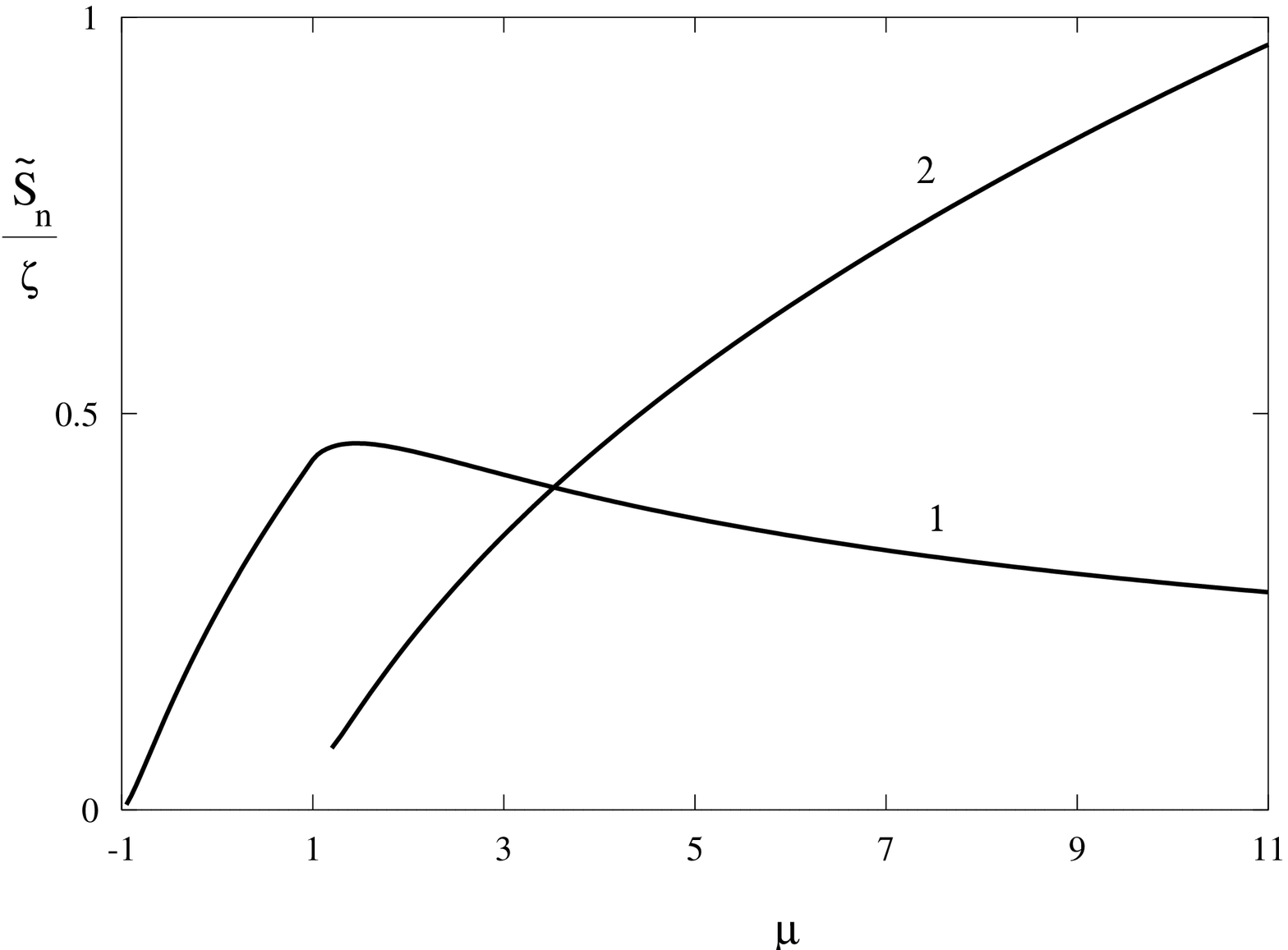}
\end{center}

\begin{figure}
\caption{The dependence of the escape activation energies $\tilde S_1=
\tilde S_2$ (line 1) and 
$\tilde S_0$ (lines 2) on the scaled frequency detuning $\mu=
\Omega/\zeta \equiv 2\omega_F[(\omega_F/2)-\omega_0]/F$ in the
limit of comparatively large fields or small damping, $\zeta\gg 1$,
for large sixth order nonlinearity, $\rho\gg 1$. }
\end{figure}

\end{document}